\documentclass[prd,showpacs,preprintnumbers]{revtex4} 

\usepackage{graphicx}
\usepackage{dcolumn}
\usepackage{bm}
\usepackage{latexsym}
\usepackage{amsfonts}
\usepackage{amssymb}
\usepackage{amsmath}
\usepackage[usenames]{color}

\newcommand{\be}{\begin{equation}}
\newcommand{\ee}{\end{equation}}
\newcommand{\bea}{\begin{eqnarray}}
\newcommand{\eea}{\end{eqnarray}}

\begin{document}
\preprint{DCPT-09/51}
\preprint{KUNS-2215}

\title{Holographic Superconductors with Higher Curvature Corrections}

\author{Ruth Gregory$^{1)}$}
\author{Sugumi Kanno$^{1)}$}
\author{Jiro Soda$^{2)}$}
\affiliation{1) Centre for Particle Theory, Durham University,
South Road, Durham, DH1 3LE, UK\\
2)Department of Physics,  Kyoto University, Kyoto, 606-8501, Japan
}

\date{\today}

\begin{abstract}
We study (3+1)-dimensional holographic superconductors in 
Einstein-Gauss-Bonnet gravity both numerically and analytically.  
It is found that higher curvature corrections make condensation harder.
We give an analytic proof of this result, and directly demonstrate
an analytic approximation method that explains the qualitative
features of superconductors as well as giving quantitatively good
numerical results. We also calculate conductivity and 
$\omega_g / T_c $, for $\omega_g$ and $T_c$ the gap in the frequency 
dependent conductivity and the critical temperature respectively. 
It turns out that the `universal' behaviour of conductivity, 
$\omega_g / T_c \simeq 8$, is not stable to the higher
curvature  corrections.  In the appendix, for completeness,
we show our analytic method can also explain 
(2+1)-dimensional superconductors.
\end{abstract}

\maketitle

\section{Introduction}

It is often felt that
the most remarkable discovery in string theory has been the AdS/CFT 
correspondence, \cite{Maldacena}, which has been further extended to the 
gauge/gravity correspondence~\cite{Benna:2008yg}. 
Interestingly, the gauge/gravity correspondence may play an important
role in condensed matter physics~\cite{Hartnoll:2009sz,Herzog:2009xv}.
In particular, the application of the gauge/gravity correspondence
to superconductors has been intensively 
studied~\cite{Gubser:2008px,HHH1,HorRob,HHH2,
Nakano:2008xc,Wen:2008pb,Albash:2008eh,Maeda:2008ir,
Maeda:2009wv,Koutsoumbas:2009pa,
Gubser:2008wv,Roberts:2008ns,Basu:2008st,Gubser:2008pf}
(see recent lecture notes \cite{Hartnoll:2009sz,Herzog:2009xv} 
for complete references). 
It would be very exciting indeed if we could explain high temperature
superconductivity from black hole physics.
In addition, from the gravity perspective the existence of scalar
condensation in black hole systems itself deserves further study in
relation to the ``no-hair'' theorems and a better understanding 
of the dressing of horizons by quantum fields \cite{Gubser:2008px,nohair}.

Remarkably, on the gauge theory side, there is a puzzle. As is well known, 
the Mermin-Wagner theorem forbids continuous symmetry breaking
in (2+1)-dimensions because of large fluctuations in lower dimensions. 
Nevertheless, holographic superconductors are found in (2+1)-dimensions.
It is possible that fluctuations in holographic superconductors are suppressed
because classical gravity corresponds to the large N limit.
If this is true, then higher curvature corrections should suppress condensation.
Of course, to examine whether or not the Mermin-Wagner theorem holds, 
we need to study 4-dimensional higher curvature gravity. 
Unfortunately, higher curvature gravity in 4 dimensions is not
particularly illuminating: higher derivative terms in general introduce
ghost degrees of freedom \cite{HDghost}, the exceptions being either 
Gauss-Bonnet or Lovelock gravity \cite{GBL}, in which specific 
combinations of the 
curvature tensors are used, or $f(R)$ gravity, \cite{fofr}, in which 
powers of the Ricci scalar only are used.
Unfortunately, the former case is non-dynamical in 4 dimensions, and
the latter case is conformally equivalent to scalar-tensor 
gravity, \cite{BWhitt}, and black hole solutions are therefore identical 
to the Einstein case \cite{BWhsoln}.

To explore this issue, we instead study 5-dimensional 
Einstein-Gauss-Bonnet gravity, which gives a known generalization
to the Schwarzschild black hole solution \cite{GBBH}.
We would also like to investigate if the universal relation 
between the gap $\omega_g$ in the frequency dependent conductivity
and the critical temperature $T_c$: $\omega_g / T_c \simeq 8$,
found in \cite{HorRob}, is stable under stringy corrections.
In the case of the quark-gluon plasma, there is a universal shear
viscosity to entropy density ratio $\eta/s =1/4\pi$~\cite{Policastro:2001yc},
and there are several analyses investigating the stability of this universal
relation~\cite{Buchel:2004di,Kats:2007mq,Myers:2008yi,Brigante:2007nu,
Brigante:2008gz,Neupane:2008dc,Ge:2008ni,Cai:2009zv} to higher curvature 
corrections. 
To the best of our knowledge, no corresponding analysis 
exists in the case of superconductors.
Hence, we look at gap frequency at a given temperature numerically to explore 
its stability under higher curvature corrections. 

To investigate the effect of the higher curvature corrections on the 
superconductor, we operate in the `probe' limit, i.e.\ where the gravitational
back reaction of the scalar and vector fields on the background geometry
is neglected. At least for temperatures near the phase transition
this should be a good approximation, and has been found to be valuable
in the Einstein limit \cite{HHH1}. 
Ideally, one would like to have a full analytic description of the 
phase transition and condensation phenomena, and in this paper
we take a modest first step in this direction.
We first prove the existence of a bound on black hole temperature above
which no condensation can occur. Since there is always an analytic solution
with vanishing scalar, \cite{HHH1}, we cannot similarly
prove the existence of a nontrivial scalar solution below $T_c$, however, a
simple matching method provides an approximate analytic solution which
explains the phase transition behaviour and gives a very good approximation
to the phase diagram.  Indeed, we can calculate the critical 
temperature analytically within a few percent in the best case. 
In a sense, this is the most important result in our paper. 
Numerical methods complete the proof of 
condensation, and are clearly necessary for fully describing the 
properties of the fields and the details of the physics.
 
The organization of the paper is as follows.
In section II, we introduce the model and numerically demonstrate the effect of
the Gauss-Bonnet term on the superconductor. We find that stringy corrections
make condensation harder. 
In section III, we present an analytic explanation of the superconductor.
We can understand the qualitative features of the superconductor
with a simple calculation. The analysis also gives fairly good
numerical results. 
In section IV, we study the conductivity and show the universality
is unstable under the stringy corrections.
We conclude in section V.
In the appendix, we present an analytic explanation
of (2+1)-dimensional superconductors for completeness. 

\section{Gauss-Bonnet superconductors}

In this section, we study the effect of Gauss-Bonnet term on the
(3+1)-dimensional superconductor using the probe limit. 
In the probe limit, gravity and matter decouple and
the system reduces to the Maxwell field and the charged scalar field
in the neutral black hole background. 

We begin with the Einstein-Gauss-Bonnet action:
\begin{eqnarray}
S=\int d^5x\sqrt{-g}\left[
R + \frac{12}{L^2} + \frac{\alpha}{2} \left(
R^{\mu\nu\lambda\rho}R_{\mu\nu\lambda\rho}
-4R^{\mu\nu}R_{\mu\nu} + R^2
\right)\right] \;,
\end{eqnarray}
where $g$ is the determinant of a metric $g_{\mu\nu}$ and
$R_{\mu\nu\lambda\rho}$, $R_{\mu\nu}$ and $R$ are the Riemann curvature tensor,
Ricci tensor, and the Ricci scalar, respectively.
We take the Gauss-Bonnet coupling constant $\alpha$ to be positive.
Here, the negative cosmological constant term $-6/L^2$ is also introduced. 
The background solution we consider is a neutral black hole~\cite{GBBH}:
\begin{eqnarray}
ds^2 = -f(r)dt^2 + \frac{dr^2}{f(r)} + \frac{r^2}{L^2}(dx^2+dy^2+dz^2)
\end{eqnarray}
where
\begin{eqnarray}
f(r) = \frac{r^2}{2\alpha}\left[
1-\sqrt{1-\frac{4\alpha}{L^2}\left(
1-\frac{M L^2}{r^4}\right)}
\right]
\label{sol}
\end{eqnarray}
Here, $M$ is a constant of integration related to the ``ADM''  mass 
of the black hole \cite{ADMass}.
The position of the horizon
defined by $f(r_H)=0$ is at $r_H=(ML^2)^{1/4}$. In order to avoid a naked
singularity, we need to restrict the parameter range as $\alpha\leq L^2/4$.
Note that in the Einstein limit ($\alpha\rightarrow 0$), the solution 
(\ref{sol}) goes
to $f(r)=\frac{r^2}{L^2}-\frac{M}{r^2}$, and $L$ can be regarded
as the curvature radius of asymptotic AdS region ($r\rightarrow\infty$).
For general $\alpha$, however, the solution (\ref{sol}) behaves as
\begin{eqnarray}
f(r)\sim\frac{r^2}{2\alpha}\left[
1-\sqrt{1-\frac{4\alpha}{L^2}}
\right]\,,
\end{eqnarray}
in the asymptotic region.
Hence, we define the effective asymptotic AdS scale by
\begin{eqnarray}\label{Leff}
L^2_{\rm eff}=\frac{2\alpha}{1-\sqrt{1-\frac{4\alpha}{L^2}}}
\to  \left\{
\begin{array}{rl}
L^2   \ , &  \quad {\rm for} \ \alpha \rightarrow 0 \\
\frac{L^2}{2}  \ , &  \quad {\rm for} \  \alpha \rightarrow \frac{L^2}{4}
\end{array}\right.
\,.
\end{eqnarray}
The Hawking temperature is given by
\begin{eqnarray}
T = \frac{1}{4\pi} f' (r)\bigg|_{r=r_H} = \frac{r_H}{\pi L^2} 
= \frac{M^{1/4}}{\pi L^{3/2}}\ ,
\end{eqnarray}
where a prime denotes derivative with respect to $r$.
This will be interpreted as the temperature of the CFT.

In this background, we
now consider a Maxwell field and a charged complex scalar field, with
the action
\begin{eqnarray}
S=\int d^5x\sqrt{-g}\left[
-\frac{1}{4}F^{\mu\nu}F_{\mu\nu}-|\nabla\psi - iA\psi|^2
-m^2|\psi|^2
\right]  \ .
\end{eqnarray}
Taking a static ansatz, $A_\mu=(\phi(r),0,0,0,0)$ and $\psi =\psi(r)$,
the equation of motion for $\phi(r)$ becomes
\begin{eqnarray}
\phi^{\prime\prime}+\frac{3}{r}\phi^\prime-\frac{2\psi^2}{f}\phi=0
\label{vector}
\end{eqnarray}
where without loss of generality $\psi$ can be taken to be
real, and satisfies
\begin{eqnarray}
\psi^{\prime\prime}+\left(
\frac{f^\prime}{f}+\frac{3}{r}\right)\psi^\prime
+\left(\frac{\phi^2}{f^2}-\frac{m^2}{f}\right)\psi=0\,.
\label{scalar}
\end{eqnarray}
Note that the Maxwell equations imply that the phase of $\psi$ 
must be constant,
which is set to zero by a residual gauge for $A_\mu$. 

We now want to solve (\ref{vector}) and (\ref{scalar}) for
the scalar and vector field. For the main part of this paper, we choose
to set the mass of the scalar field to be $m^2=-3/L^2$, so that
the mass remains the same as we vary $\alpha$. Note however, that
because of the variation of the effective asymptotic AdS curvature,
(\ref{Leff}), with $\alpha$ relative to $L$ means that this mass
actually {\it increases} (i.e.\ becomes less negative) with respect
to the asymptotic AdS scale. On the other hand, while setting
$m^2 = -3/L_{\rm eff}^2$ has the advantage of fixing the mass
relative to the asymptotic AdS scale, this mass now varies with 
respect to the physical measurables of black hole mass and temperature
as we vary $\alpha$. Since condensation is a temperature dependent 
phenomenon, we believe that fixing the scalar mass with respect
to the black hole is the correct physical choice, however, we have
also checked that for the alternative choice of mass the same
qualitative features occur as we vary $\alpha$. 

In order to solve our equations we need to impose regularity at
the horizon and the AdS boundary:

\noindent $\bullet$ Regularity at the horizon gives two conditions:
\begin{eqnarray}
\phi(r_H)=0,\hspace{1cm}
\psi(r_H)=-\frac{4}{3}r_H\psi^\prime(r_H) \ .
\end{eqnarray}

\noindent $\bullet$ Asymptotically ($r\rightarrow\infty$) the 
solutions are found to be:
\begin{eqnarray}
\phi(r)=\mu - \frac{\rho}{r^2}\,,\hspace{0.5cm} 
\psi=\frac{C_{-}}{r^{\lambda_-}}+\frac{C_{+}}{r^{\lambda_+}}\,,
\label{r:boundary}
\end{eqnarray}
where $\lambda_\pm=2\pm\sqrt{4-3\left(\frac{L_{\rm eff}}{L}\right)^2}$. 
Here, $\mu$ and $\rho$ are interpreted as a chemical
potential and charge density, respectively. Note that these are
not entirely free parameters, as there is a scaling degree of 
freedom in the equations of motion. As in \cite{HHH1}, we impose 
that $\rho$ is fixed, which determines the scale of this system. 
For $\psi$, both of these falloffs are normalizable, so we can 
impose the condition either $C_{-}$ or $C_{+}$ vanish. 
We take $C_{-}=0$,  for simplicity.

\begin{figure}[ht]
\includegraphics[height=7cm, width=9.5cm]{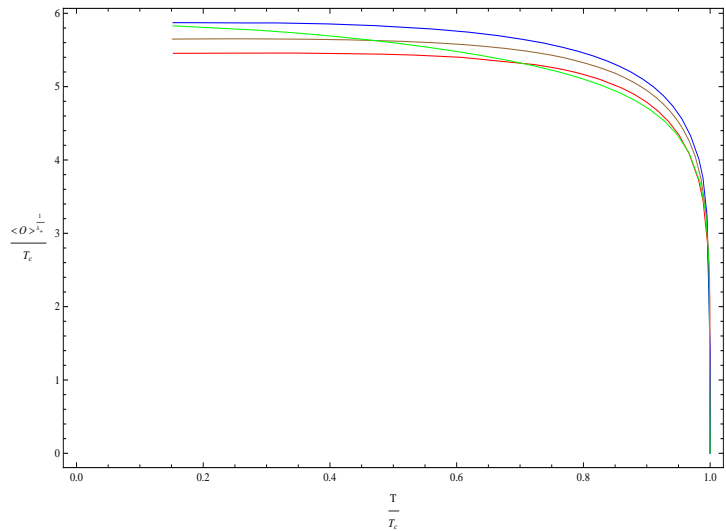}
\caption{The condensate as a function of temperature for various values
of $\alpha$. The (lowest) red line is for $\alpha=0.0001$, the middle
brown plot is $\alpha=0.1$, the top blue line is $\alpha=0.2$ and
the remaining line intersecting the other three in green is 
the Chern-Simons limit $\alpha=0.25$. Note that while the generic
Einstein-Gauss-Bonnet behaviour is to level out for $T \leq T_c/2$,
the Chern-Simons limit has a much stronger variation of the
condensate with temperature.}
\label{fig:cond}
\end{figure}

According to the AdS/CFT correspondence, we can interpret 
$ \langle {\cal O} \rangle \equiv C_{+}$, 
where ${\cal O}$ is the operator dual to the scalar field.
Thus, we are going to calculate the condensate $\langle {\cal O} \rangle$ 
for fixed charge density.
The results are shown in Figure \ref{fig:cond}. 
From Fig.\ \ref{fig:cond}, we see the GB term
makes the condensation gap larger. 
We also see that the Chern-Simons limit shows a slightly different
dependence of the condensate on temperature. 
This can be understood from the behaviour of gravity near the horizon.
In the Chern-Simons limit, $\alpha =L^2/4$, we get
\begin{eqnarray}
  f(r)= \frac{2r^2}{L^2} \left(1- \frac{\sqrt{ML^2}}{r^2}\right) \ .
\end{eqnarray}
Hence, the correction to the AdS quadratic gravitational potential 
dependence is simply a constant instead of a $1/r^2$ dependence, leading
to more gentle tidal behaviour. The process of scalar condensation
(or the formation of scalar hair) can be understood as arising in part
from the `negative' mass of the scalar field, but also as arising from
the potential well that occurs near the horizon. For black holes with large 
mass, this well is too broad and shallow to allow for the formation of
a nonzero scalar, however, for small black hole mass, the strong
curvature near the horizon is amenable to condensation. (Refer to the analytic
arguments in section III which show how the behaviour of the gravitational
potential interacts with features of the scalar condensate.) At some 
stage further decreasing the mass of the black hole does not alter the 
shape of the condensate much, as the scalar is already sampling
regions of strong curvature. However, the CS limit has a rather 
different and smoother profile near the horizon, therefore it is 
not surprising that decreasing the black hole mass
in this case has more impact on the details of the scalar field.

Numerically, we found that increasing $\alpha$ resulted in a decrease
of the critical temperature: $T_c = 0.198 \rho^{1/3}$ for $\alpha =0.0001$,
$T_c = 0.186 \rho^{1/3}$ for $\alpha =0.1$, 
$T_c = 0.171 \rho^{1/3}$ for $\alpha =0.2$, and $T_c = 0.158 \rho^{1/3}$ 
for $\alpha =0.25$ (see also figure \ref{fig:bound}).
Thus the effect of $\alpha$ is to make it harder for scalar hair to 
form. Changing the
scalar mass to $m^2 = -3/L^2_{\rm eff}$ gives a similar, though
less marked, behaviour, for example $T_c = 0.181 \rho^{1/3}$ 
for $\alpha = 0.2$.
We can therefore conclude, as expected, that
the higher curvature corrections make it harder for the
scalar hair to form. One can expect this tendency
to be the same even in (2+1)-dimensions,
however, it remains obscure to what extent this suppression
affects the physics of holographic superconductors in (2+1)-dimensions.

We have thus numerically verified that Gauss-Bonnet superconductors exist.
However, we would ideally like to have an analytic understanding of 
condensation to back up this numerical work. This is what we now 
turn to.

\section{Superconductors in a nutshell}

Although in the previous section we used numerical integration
to explicitly demonstrate the condensation phenomenon, ideally
we would like to obtain an analytic understanding in parallel.
Since our equations are nonlinear and coupled, we cannot
derive analytic solutions in closed form, however we can deduce a 
great deal of information analytically.
We first prove the nonexistence of condensation for large $T$
before explicitly deriving the phase diagram analytically by 
using approximate solutions.

Note that the trivial solution to (\ref{vector}) and (\ref{scalar})
\begin{eqnarray}
\phi = \phi_0(r) &=& \frac{\rho}{r_H^2} \left ( 1 - \frac{r_H^2}{r^2} \right ) \\
\psi & \equiv & 0
\end{eqnarray}
always exists. We will now prove that there is a temperature above
which this is the only solution.

First consider the $\phi$ equation (\ref{vector}). Let
$\phi(r) = \phi_0(r) + \delta \phi$, where $\phi_0(r)$
is defined above. Then (\ref{vector}) implies
\be
\left ( r^3 \delta\phi'\right )' \geq 0
\ee
however, as $r\to\infty$, $r^3 \phi' \to 2\rho = r^3\phi_0'$, hence
$r^3 \delta\phi' \to 0$ at infinity, and using $\delta\phi = 0$ at $r_H$
we have that $ \delta \phi' \leq 0 $.
Hence 
\be
\phi(r) \leq \phi_0(r) \;.
\ee

\begin{figure}[ht]
\includegraphics[height=7cm, width=9.5cm]{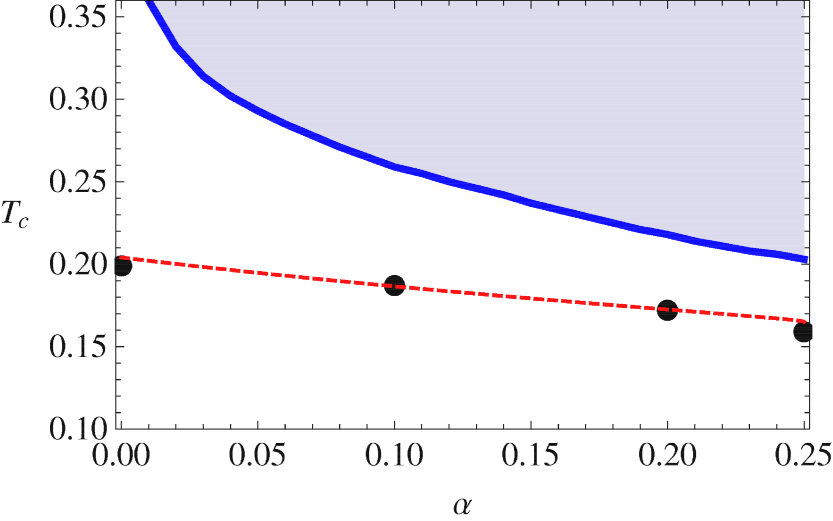}
\caption{A comparison of analytic and numerical results. The shaded
region is that in which the geometry forbids the possibility of a scalar
condensate from (\ref{anbnd}). The dashed line indicates the analytic
approximation of the value of $T_c$ obtained by matching methods, 
(\ref{mainTc}). The data points are the exact numerical results. 
For simplicity $\rho$ and $L$ have been set to 1.}
\label{fig:bound}
\end{figure}

Next consider the scalar field, and define the variable $X=r\psi$:
\be
X'' + \left ( \frac{f'}{f} + \frac{1}{r} \right ) X' + \left (
\frac{\phi^2}{f^2} + \frac{3}{L^2f} - \frac{f'}{rf} - \frac{1}{r^2}
\right ) X = 0
\ee
Now, the boundary conditions at the horizon imply $X'_H = X_H/4r_H$,
and at infinity, $rfX' \to 0$, thus the existence of a condensate
requires a turning point in $X$, $X'(r_{_T})=0$, with $X''<0$ for $X>0$. 
This in turn requires 
\be\label{anbnd}
\frac{\phi_0^2(r_{_T})}{f(r_{_T})} + \frac{3}{L^2} - \frac{f'(r_{_T})}{r_{_T}} 
- \frac{f(r_{_T})}{r_{_T}^2} > \frac{\phi^2(r_{_T})}{f(r_{_T})} + \frac{3}{L^2} 
- \frac{f'(r_{_T})}{r_{_T}} - \frac{f(r_{_T})}{r_{_T}^2} > 0
\ee
at the turning point. By inputting the form of $\phi_0(r)$, 
it is easy to see that if $M$ is too large, this inequality can
never be satisfied, as the combination of $\phi$ and the geometry to 
the LHS of (\ref{anbnd}) is always negative. 
This gives a loose upper bound on the critical
temperature as shown in figure \ref{fig:bound}. (For $\alpha =0$ we
need to use the fact that $\int_{r_{_T}}^\infty (rfX') =0$ to bound $M$.
This also gives a tighter analytical bound for nonzero $\alpha$, however,
the above argument is more direct.)

We have thus numerically verified that Gauss-Bonnet superconductors exist.
Having shown that there is a critical temperature below which there
is no barrier to condensation, we will now show we can understand the 
essential features of
condensation by using approximation techniques. 

Once again, let us change variables and set $z=\frac{r_H}{r}$. Under this
transformation equations (\ref{vector}) and (\ref{scalar}) become
\begin{eqnarray}
&&\phi^{\prime\prime}-\frac{1}{z}\phi^\prime
-\frac{r_H^2}{z^4}\frac{2\psi^2}{f}\phi=0
\label{vector3}\\
&&\psi^{\prime\prime}+\left(\frac{f^\prime}{f}-\frac{1}{z}\right)\psi^\prime
+\frac{r_H^2}{z^4}\left(\frac{\phi^2}{f^2}+\frac{3}{L^2f}\right)
\psi=0
\label{scalar3}
\end{eqnarray}
where a prime now denotes $\frac{d}{dz}$. 
The region $r_H<r<\infty$ now corresponds to $0<z<1$.
The boundary conditions now become:

\noindent $\bullet$ Regularity at the horizon $z=1$ gives
\begin{eqnarray}
\phi(1)=0\,,\hspace{1cm}\psi^\prime(1)=\frac{3}{4}\psi(1)\,.
\label{regularity2}
\end{eqnarray}

\noindent $\bullet$ In the asymptotic AdS region: $z\rightarrow 0$, 
the solutions are
\begin{eqnarray}
\phi=\mu - qz^2\,,
\hspace{1.5cm}
\psi=D_-z^{\lambda_-} + D_+z^{\lambda_+}\,,
\label{boundary2}
\end{eqnarray}
where $\lambda_\pm$ is the same as in equation (\ref{r:boundary}). 
As boundary conditions, we fix $qr_H^2$ and take $D_-$ to be zero.

We now find leading order solutions near the horizon and asymptotically,
say $1 \geq z >z_m $ and $ z_m > z \geq 0$, and then match these smoothly
at the intermediate point, $z_m$. As a consequence, we will demonstrate
the phase transition phenomenon directly, and derive an (approximate) 
analytic expression for the critical temperature.
Moreover, we will have a much better analytical understanding of
$\alpha$ dependence of the critical temperature, as the proof above
only gives a loose bound on the critical temperature and only indirect
access to an expression.

\subsection{Solution near the horizon: $z=1$}

We can expand $\phi$ and $\psi$ in a Taylor series near the horizon as:
\begin{eqnarray}
\phi(z)&=&\phi(1)-\phi^\prime(1)(1-z)+\frac{1}{2}\phi^{\prime\prime}(1)(1-z)^2
+\cdots\\
\psi(z)&=&\psi(1)-\psi^\prime(1)(1-z)+\frac{1}{2}\psi^{\prime\prime}(1)(1-z)^2
+\cdots
\end{eqnarray}
From (\ref{regularity2}), we have $\phi(1)=0$ and 
$\psi^\prime(1)=\frac{3}{4}\psi(1)$, and without loss of generality
we take $\phi^\prime(1)<0$, $\psi(1)>0$ to have $\phi(z)$ and $\psi(z)$ 
positive. Expanding (\ref{vector3}) near $z=1$ gives:
\begin{eqnarray}
\phi^{\prime\prime} (1) &=&\frac{1}{z}\phi^\prime~\big|_{z=1}
+\frac{r_H^2}{z^4}\frac{2\psi^2}{f}\phi~\bigg|_{z=1}
\nonumber\\
&=&\phi^\prime(1)-\frac{2r_H^2\psi(1)^2}{z^4(1-z)f'(1)}
\left(-\phi^\prime(1)(1-z)+\frac{1}{2}\phi^{\prime\prime}(1)(1-z)^2
+\cdots\right)~\bigg|_{z=1}\nonumber\\
&=&\left(1-\frac{L^2}{2}\psi(1)^2\right)\phi^\prime(1)
\end{eqnarray}
Thus, we get the approximate solution
\begin{eqnarray}
\phi(z)=-\phi^\prime(1)(1-z)+\frac{1}{2}\left(1-\frac{L^2}{2}\psi(1)^2\right)
\phi^{\prime}(1)(1-z)^2
+\cdots
\label{phi:horizon2}
\end{eqnarray}

Similarly, from (\ref{scalar3}), the 2nd order coefficients 
of $\psi$ can be calculated as
\begin{eqnarray}
\psi^{\prime\prime}(1)
&=&\frac{1}{z}\psi^\prime~\bigg|_{z=1}
-\frac{z^4f^\prime\psi^\prime+3\frac{r_H^2}{L^2}\psi}{z^4f}~\bigg|_{z=1}
-\frac{r_H^2\phi^2}{z^4f^2}\psi~\bigg|_{z=1}
\nonumber\\
&=&\psi^\prime(1)
-\frac{4z^3f^\prime\psi^\prime+z^4f^{\prime\prime}\psi^\prime
+z^4f^\prime\psi^{\prime\prime}
+3\frac{r_H^2}{L^2}\psi^\prime}{4z^3f+z^4f^\prime}~\bigg|_{z=1}
-\frac{r_H^2\psi \left( -\phi^\prime(1)(1-z)+\cdots\right)^2}
{f'(1)^2 (1-z)^2} ~\bigg|_{z=1}
\nonumber\\
&=&-\frac{5}{4}\psi^\prime(1)+8\frac{\alpha}{L^2}\psi^\prime(1)
-\psi^{\prime\prime}(1)-\frac{L^4}{16r_H^2}\phi^\prime(1)^2\psi(1)
\end{eqnarray}
where we used l'H$\hat{\rm o}$pital's rule at the second term in the second 
line. Thus, we get
\begin{eqnarray}
\psi^{\prime\prime}(1)=\left(-\frac{5}{8}+\frac{4\alpha}{L^2}
\right)\psi^\prime(1)
-\frac{L^4}{32r_H^2}\phi^\prime(1)^2\psi(1)
\end{eqnarray}
After eliminating $\psi^\prime(1)$ from above equation 
by using Eq.~(\ref{regularity2}), we find
an approximate solution near the horizon as
\begin{eqnarray}
\psi(z)=\frac{1}{4}\psi(1)+\frac{3}{4}\psi(1)z
+\left(-\frac{15}{64}+\frac{3\alpha}{2L^2}
-\frac{L^4}{64r_H^2}\phi^{\prime}(1)^2\right)\psi(1)(1-z)^2
+\cdots
\label{psi:horizon2}
\end{eqnarray}

\subsection{Solution near the asymptotic AdS region: $z=0$}

From (\ref{boundary2}), $\phi$ and $\psi$ in the asymptotic region
are given by
\begin{eqnarray}
\phi(z)=\mu-qz^2\,,\hspace{1cm}
\psi(z)=D_+z^{\lambda_+}
\label{sol:infty2}
\end{eqnarray}
where $qr_H^2$ is fixed and we have set $D_-=0$ from the boundary condition.

\subsection{Matching and Phase Transition}

Now we will match the solutions (\ref{phi:horizon2}), (\ref{psi:horizon2})
and (\ref{sol:infty2}) at $z_m$. 
Interestingly, allowing $z_m$ to be arbitrary does not change 
qualitative features of the analytic approximation, more importantly,
it does not give a big difference in numerical values, therefore for
simplicity in demonstrating our argument we will take $z_m =1/2$. 
In order to connect our two asymptotic solutions smoothly, 
we require the following 4 conditions: 
\begin{eqnarray}
&&\mu-\frac{1}{4}q=\frac{1}{2}b-\frac{1}{8}b\left(1-\frac{L^2}{2}a^2\right)
\label{phi}\\
&&-q=-b+\frac{1}{2}b\left(1-\frac{L^2}{2}a^2\right)
\label{dphi}\\
&&D_+\left(\frac{1}{2}\right)^{\lambda_+}=\frac{5}{8}a+\frac{1}{4}a
\left(-\frac{15}{64}+\frac{3\alpha}{2L^2}-\frac{L^4}{64r_H^2}b^2\right)
\label{psi}\\
&&2\lambda_+D_+\left(\frac{1}{2}\right)^{\lambda_+}=\frac{3}{4}a
-a\left(-\frac{15}{64}+\frac{3\alpha}{2L^2}-\frac{L^4}{64r_H^2}b^2
\right)
\label{dpsi}
\end{eqnarray}
where we have set $\psi(1)\equiv a$ and $-\phi^\prime(1)\equiv b\,\,(a,b>0)$
for clarity.  Now, the AdS/CFT dictionary gives a relation
$\langle {\cal O} \rangle\equiv L D_{+} r_H^{\lambda_+} L^{-2\lambda_{+}}$,
hence we need to compute $D_{+}$. 
From (\ref{psi}) and (\ref{dpsi}) we obtain
\begin{eqnarray}
D_+=\frac{13}{8}\frac{2^{\lambda_+}}{\lambda_++2}\; a \;\;.
\label{D}
\end{eqnarray}
Using (\ref{phi}) and (\ref{dphi}), $a$ is expressed by 
\begin{eqnarray}
a^2=\frac{4q}{L^2b}\left(1-\frac{b}{2q}\right) \;\;,
\label{a}
\end{eqnarray}
where $b$ is obtained from (\ref{psi}) and (\ref{dpsi}) assuming
$a\neq 0$ (i.e.\ the scalar solution is non-trivial) as:
\begin{eqnarray}
b=8\frac{r_H}{L^2} \sqrt{\frac{5\lambda_+-3}
{2(\lambda_++2)}-\frac{15}{64}+\frac{3\alpha}{2L^2}} \;\;.
\label{b}
\end{eqnarray}

Now we go back to the original variable, $r$, and compare the results with 
those in \cite{HorRob}. First of all, we should note the relation
$\rho=q~r_H^2$. We also define $\tilde{b}$ by $b=\tilde{b}r_H /L^2$.
Using the Hawking temperature $T=\frac{r_H}{\pi L^2}$, we can 
rewrite (\ref{a}) as
\begin{eqnarray}\label{adefn}
a^2=\frac{2}{L^2}\frac{T_c^3}{T^3}\left(1-\frac{T^3}{T_c^3}\right)\;\;,
\end{eqnarray}
where we have defined $T_c$ as
\begin{eqnarray}
T_c=\left(\frac{2\rho}{{\tilde b} L}  \right)^{1/3}\frac{1}{\pi L}\;\;.
\label{mainTc}
\end{eqnarray}

We can now read off the expectation value  $\langle {\cal O} \rangle$ from
(\ref{D}) and (\ref{adefn}) as:
\begin{eqnarray}
\frac{\langle {\cal O} \rangle^{\frac{1}{\lambda_+}}}{T_c}
=2\pi \left(\frac{13}{8}\frac{\sqrt{2}}{\lambda_++2}
\right)^{\frac{1}{\lambda_+}}\frac{T}{T_c}
\left[~\frac{T_c^3}{T^3}\left(1-\frac{T^3}{T_c^3}\right)
~\right]^{\frac{1}{2\lambda_+}}  \ ,
\end{eqnarray}
where we have normalized by the critical temperature to obtain
a dimensionless quantity.  We find that $\langle {\cal O} \rangle$ 
is zero at $T=T_c$, the critical point, and
condensation occurs for $T<T_c$. We also see
a behaviour $\langle {\cal O} \rangle \propto(1-T/T_c)^{1/2}$
which is a typical mean field theory result 
for a the second order phase transition. 

Next, we evaluate the critical temperature from (\ref{mainTc}).
The value of $T_c$ is $0.201\rho^{1/3}/L$ 
when the Gauss-Bonnet term is absent, this should be compared with
the numerical result $T_c=0.198\rho^{1/3}/L$ in \cite{HorRob}. 
We therefore see that our analytic approximation is good. 
Moreover, as $\alpha$ increases to $0.1, 0.2$ and $0.25$,
$T_c$ decreases to $0.196, 0.191$ and $0.188$ respectively,
which is in good agreement with our numerical results.

Thus, we have (approximately) reproduced our numerical results 
from a simple analytic calculation. In particular, we have calculated 
extremely good estimates of the critical temperatures, and
revealed how the structure of the interaction term 
has produced the phase transition.

\section{Conductivity and universality}

We now calculate the conductivity, $\sigma$, of our boundary theory.
In \cite{HorRob}, the conductivity for various cases was calculated and
it was found that there is a universal relation
\begin{eqnarray}
\frac{\omega_g}{T_c} \simeq 8 \ ,
\end{eqnarray}
with deviations of less than 8 \%. The purpose of this section is to
examine if this universality holds in the presence of stringy corrections.

As $A_\mu$ in the bulk corresponds to the four-current $J_\mu$ on 
the CFT boundary, we can calculate the conductivity by considering
perturbation of $A_\mu$. The spatial
components of $A_\mu$ are decomposed into longitudinal and
transverse modes: $A_i=(\partial_i\chi, A_i^\perp)$. These
linearized perturbations are decoupled from each other and can be
studied separately. The linearized equation of motion for
$A_i^\perp(t,r,x^i)=A(r)e^{i{\bf k}\cdot{\bf x}-i\omega t}e_i$,
which corresponds to the current density, is 
\begin{eqnarray}
A^{\prime\prime}+\left(\frac{f^\prime}{f}+\frac{1}{r}\right)A^\prime
+\left(\frac{\omega^2}{f^2}-\frac{\bf{k}^2}{r^2f}-\frac{2}{f}\psi^2\right)A
=0 \; .
\label{A:eq}
\end{eqnarray}
We solve this under the following boundary conditions near the horizon:
\begin{eqnarray}
A(r) \sim f(r)^{-i\frac{\omega}{4r_H}} \ ,
\label{ingoing}
\end{eqnarray}
which corresponds to no outgoing radiation at the horizon. In
the asymptotic AdS region $(r\rightarrow\infty)$, 
the general solution takes the form
\begin{eqnarray} \label{genasoln}
A=A^{(0)}+\frac{A^{(2)}}{r^2}
+\frac{A^{(0)}(\omega^2-{\bf{k}}^2)L_{\rm eff}^2}{2}
\frac{\log\Lambda r}{r^2}
\end{eqnarray}
where $A^{(0)}$, $A^{(2)}$ and $\Lambda$ are arbitrary integration 
constants. Note the appearance of the arbitrary scale $\Lambda$, which
leads to a logarithmic divergence in the Green's function, as explained
in \cite{HorRob}. Since this can be removed by an appropriate 
boundary counterterm, this scale will disappear from the results.  

From linear response theory, the conductivity can be calculated
by the formula 
\begin{eqnarray}
  \sigma (\omega) = \frac{1}{i \omega} G^{R}(\omega , k=0) \ ,
\end{eqnarray}
where $k$ is the wavenumber. The retarded Green function $G^R$ can be
calculated through the AdS/CFT correspondence~\cite{Son:2002sd} as:
\begin{eqnarray}
G^R = - \lim_{r \rightarrow \infty}
f(r) r A A'  \ . 
\end{eqnarray}
Thus, by using the solution (\ref{genasoln}), the conductivity is given by
\begin{eqnarray}
\sigma=\frac{2A^{(2)}}{i\omega A^{(0)}}\bigg|_{\bf{k}=0}+\frac{i\omega}{2} \ .
\end{eqnarray}
We therefore need to solve (\ref{A:eq}) numerically with the boundary condition
(\ref{ingoing}) to obtain $A^{(0)}$ and $A^{(2)}$ asymptotically.

The plots in figures 3$-$6 show
the results of this numerical integration for $\alpha=0.0001, 0.1, 0.2$ 
and $0.25$ at temperatures $T/T_c\approx 0.152, 0.151, 0.152$ 
and $0.152$, respectively. The red line represents
the real part, and blue line the imaginary part of $\sigma$.
Taking look at the imaginary part of the conductivity, we see
a pole exists at $\omega=0$. From the Kramers-Kronig relations,
this implies the real part of the conductivity contains a delta function.

\begin{figure}[ht]
\begin{center}
\begin{minipage}{8cm} 
\begin{center}
\includegraphics[width=8cm]{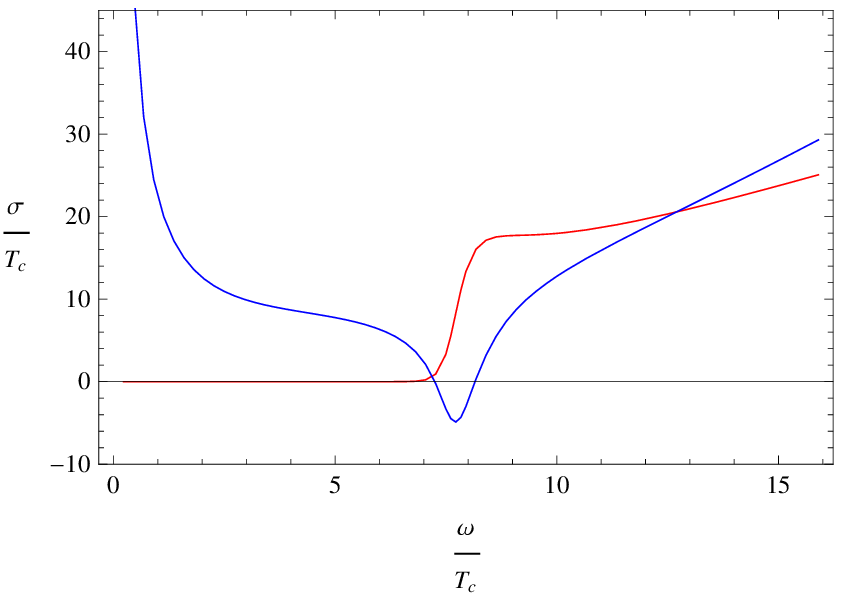}\vspace{0cm}
\caption{Conductivity for $\alpha=0.0001$ case}
\end{center}
\end{minipage}
\hspace{1mm}
\begin{minipage}{8cm}
\begin{center}\vspace{0cm}
\includegraphics[width=8cm]{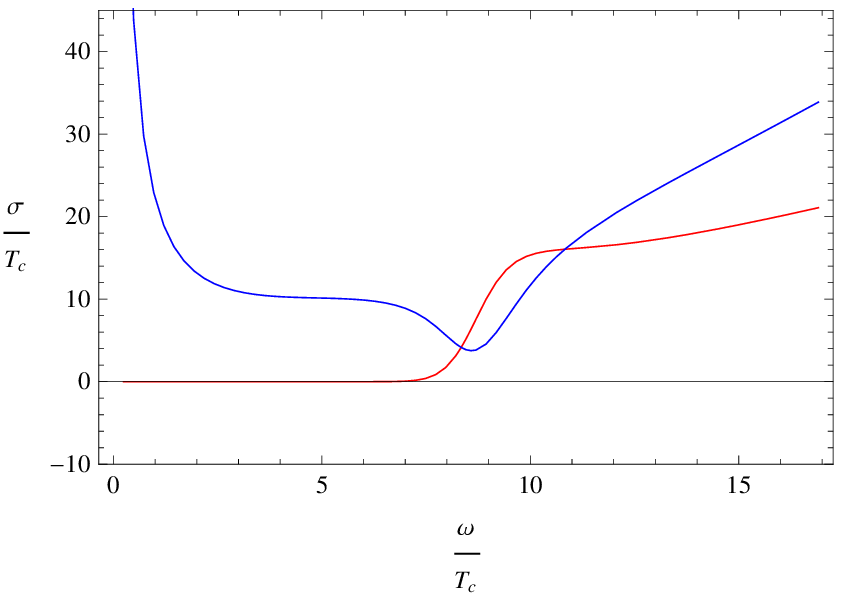}\vspace{0cm}
\caption{Conductivity for $\alpha=0.1$ case}\end{center}
\end{minipage}
\vskip 1cm
\begin{minipage}{8cm} 
\begin{center}
\includegraphics[width=8cm]{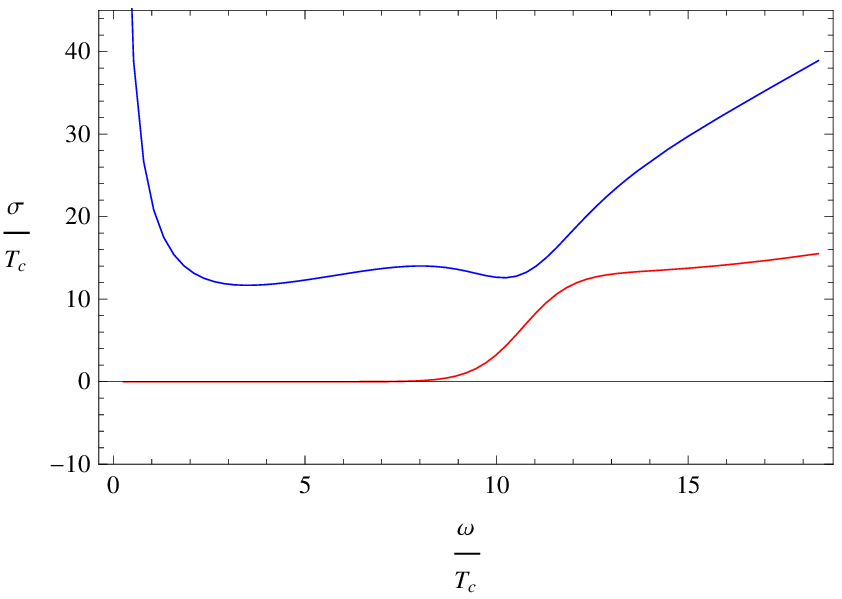}\vspace{0cm}
\caption{Conductivity for $\alpha=0.2$ case}
\end{center}
\end{minipage}
\hspace{1mm}
\begin{minipage}{8cm} 
\begin{center}
\includegraphics[width=8cm]{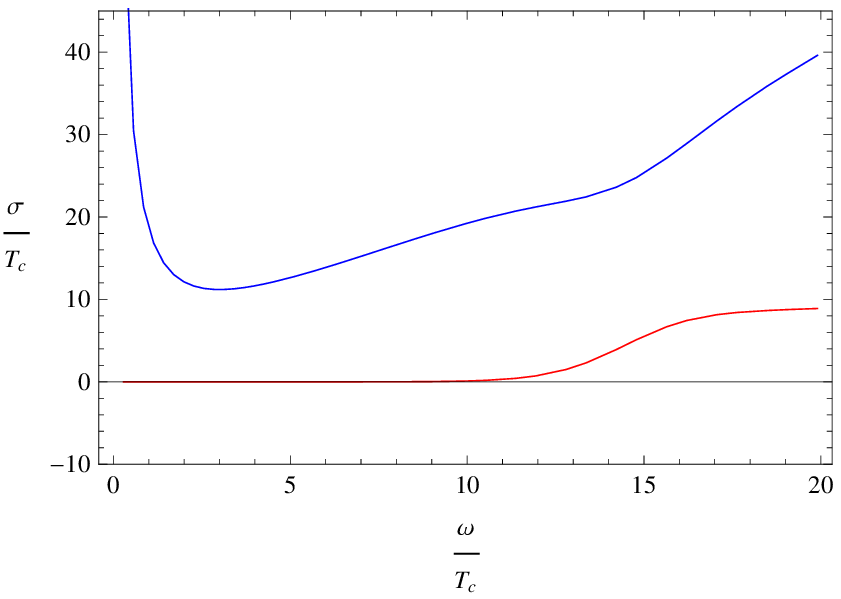}\vspace{0cm}
\caption{Conductivity for $\alpha=0.25$ case}
\end{center}
\end{minipage}
\hspace{1mm}
\end{center}
\end{figure}

Clearly, the real part of the conductivity shows a frequency gap
which indicates a gap in the spectrum of charged excitations.
As $\alpha$ increases, the gap frequency (normalized by $T_c$) becomes
large. As we noticed that condensation is an increasing function of $\alpha$
this tendency is consistent with the conventional relation
$\omega_g \propto \ \langle {\cal O} \rangle$. 
We see the universal relation $\frac{\omega_g}{T_c}\approx 8$ 
found in \cite{HorRob} is unstable in the presence of the Gauss-Bonnet 
correction. We have also checked that this conclusion is not affected 
by choosing the alternative scalar mass, $M^2 = -3/L_{\rm eff}^2$.

\section{Conclusion}

We have studied holographic superconductors 
in the presence of Gauss-Bonnet corrections to the 
gravitational action. 
Motivated by the Mermin-Wagner theorem, we have investigated if
the higher derivative corrections suppress the phase transition or not.
We numerically solved the system in the probe limit and obtained
phase diagrams for various Gauss-Bonnet couplings $\alpha$, and calculated
the critical temperatures. As we increase $\alpha$, the critical temperature 
decreases, thus it turns out that stringy corrections make condensation 
harder. However, we did not reach the point that the critical temperature 
of the transition vanishes for changing $\alpha$.
We would also expect this to apply to the (2+1)-dimensional case,
however, it is beyond the scope of this paper to determine
if this could destroy holographic superconductors in (2+1)-dimensions.

To understand phase transition phenomena, we also conducted an analytic
analysis of the coupled nonlinear equations, finding an approximate 
analytic solution. In spite of the apparent crudity of this approximation, 
we have analytically demonstrated the phase transition. 
Surprisingly, it turned out that the analytical
method gave good agreement with the numerical results. 
In particular, we have calculated the critical temperature analytically.
We obtained $T_c = 0.201 \rho^{1/3}$, which is close to
the numerical result $T_c = 0.198 \rho^{1/3}$ for $\alpha \to 0$.
We also applied the same method to the (2+1)-dimensional superconductor,
presented in the appendix.
The resultant critical temperature was $T_c =0.103 \sqrt{\rho}$,
which should be compared with the numerical result $T_c = 0.118 \sqrt{\rho}$
\cite{HHH1}.

Our other purpose was to examine the universality of
the gap frequency to the critical temperature ratio. 
By calculating conductivity, we found the universal behaviour of
conductivity $\omega_g / T_c \simeq 8$ was unstable to the stringy
corrections.  
 
There are many issues to be investigated further. The obvious 
next step is to incorporate back reaction, which is particularly important in
the low temperature regime which corresponds to small black holes.
In that case, the stability of black holes should be 
considered~\cite{Dotti:2004sh,Dotti:2005sq,Gleiser:2005ra,Takahashi:2009dz}.
Although we have investigated the stability of the superconductor
under stringy corrections, it is also intriguing to study
the dynamical stability of the condensation phase, as well as
other aspects of 
superconductors~\cite{Kim:2009kb,Albash:2009iq,Montull:2009fe,KM}. 

\acknowledgements
{
We wish to thank Pau Figueras, Takashi Okamura,
and Paul Sutcliffe for useful discussions. 
SK is supported by an STFC rolling grant.
JS is supported by the Japan-U.K. Research Cooperative Program, 
Grant-in-Aid for  Scientific Research Fund of the Ministry of 
Education, Science and Culture of Japan No.18540262
and by the Grant-in-Aid for the Global COE Program 
``The Next Generation of Physics, Spun from Universality and Emergence". 
}

\appendix

\section{ Analytical Approach to (2+1)-dimensional superconductors}

In the main text, we have shown a simple analytic treatment gives
a good explanation of superconductivity. Here for completeness, 
we show (2+1)-dimensional
superconductors can be also explained using the same method. 

In the 4-dimensional case, we have the following equations
\begin{eqnarray}
&& \phi^{\prime\prime}+\frac{2}{r}\phi^\prime-\frac{2\psi^2}{f}\phi=0 \ ,
\label{vector:4} \\
&& \psi^{\prime\prime}+\left(
\frac{f^\prime}{f}+\frac{2}{r}\right)\psi^\prime
+\left(\frac{\phi^2}{f^2}-\frac{m^2}{f}\right)\psi=0\,,
\label{scalar:4}
\end{eqnarray}
where now
\be
f(r) = \frac{r^2}{L^2} \left ( 1 - \frac{r_H^3}{r^3} \right )
\ee
with $r_H = (ML^2)^{1/3}$.
We set the mass of the scalar field, $m^2=-2/L^2$, as in \cite{HHH1}.
By changing to the $z$ variable as before, $z=\frac{r_H}{r}$, 
(\ref{vector:4}) and (\ref{scalar:4}) become
\begin{eqnarray}
&&\phi^{\prime\prime}-\frac{2L^2\psi^2}{z^2(1-z^3)}\phi=0
\label{vector2}\\
&&\psi^{\prime\prime}-\frac{2+z^3}{z(1-z^3)}\psi^\prime
+\left(\frac{L^4\phi^2}{r_H^2(1-z^3)^2}+\frac{2}{z^2(1-z^3)}
\right)\psi=0
\label{scalar2}
\end{eqnarray}
where a prime now denotes $\frac{d}{dz}$. 
Next we consider the boundary conditions with these new variables. 
Regularity at the horizon, $z=1$, requires
\begin{eqnarray}
\phi(1)=0\,,\hspace{1cm}\psi^\prime(1)=\frac{2}{3}\psi(1)\,,
\label{regularity}
\end{eqnarray}
and the asymptotic solution in the AdS region, $z\rightarrow 0$, reads
\begin{eqnarray}
\phi=\mu - qz\,,
\hspace{1.5cm}
\psi=C_1z + C_2z^2 \ .
\label{boundary}
\end{eqnarray}
As in \cite{HHH1} we fix the charge $qr_H$ and take $C_1$ to be zero.

We now find an approximate solution around both $z=1$ and $z=0$ 
using Taylor expansion as before, then connect these solutions 
between $z=1$ and $z=0$.

\subsection{Solution near the horizon: $z=1$}
We expand $\phi$ and $\psi$ as
\begin{eqnarray}
\phi(z)&=&\phi(1)-\phi^\prime(1)(1-z)+\frac{1}{2}\phi^{\prime\prime}(1)(1-z)^2
+\cdots\\
\psi(z)&=&\psi(1)-\psi^\prime(1)(1-z)+\frac{1}{2}\psi^{\prime\prime}(1)(1-z)^2
+\cdots
\end{eqnarray}
From the boundary condition (\ref{regularity}), $\phi(1)=0$ and $\psi^\prime(1)=\frac{2}{3}\psi(1)$, and we again set $\phi^\prime(1)<0$ and $\psi(1)>0$
for positivity of $\phi(z)$ and $\psi(z)$.

First we compute the 2$^{\rm nd}$ order coefficient $\phi$ using 
(\ref{vector2}) as
\begin{eqnarray}
\phi^{\prime\prime}~\big|_{z=1}
=\frac{2L^2\psi^2}{z^2(1-z^3)}\phi~\bigg|_{z=1}
= -\frac{2}{3}L^2\phi^\prime(1)\psi(1)^2 > 0 \; ,
\end{eqnarray}
giving
\begin{eqnarray}
\phi(z)=-\phi^\prime(1)(1-z)-\frac{1}{3}L^2\psi(1)^2\phi^{\prime}(1)(1-z)^2
+\cdots \; .
\label{phi:horizon}
\end{eqnarray}

The 2nd derivative of $\psi$ is calculated similarly as
\begin{eqnarray}
\psi^{\prime\prime}~\big|_{z=1}
&=&\frac{2+z^3}{z(1-z^3)}\psi^\prime~\bigg|_{z=1}
-\frac{2}{z^2(1-z^3)}\psi~\bigg|_{z=1}
-\frac{L^4\phi^2}{r_H^2(1-z^3)^2}\psi~\bigg|_{z=1}
\nonumber\\
&=&\frac{(z^4+2z)\psi^{\prime\prime}+4z^3\psi^\prime}{2z-5z^4}~\bigg|_{z=1}
-\frac{L^4}{9r_H^2}\phi^\prime(1)^2\psi(1)
\nonumber\\
&=&-\psi^{\prime\prime}(1)-\frac{4}{3}\psi^\prime(1)
-\frac{L^4}{9r_H^2}\phi^\prime(1)^2\psi(1)
\end{eqnarray}
Thus
\begin{eqnarray}
\psi^{\prime\prime}(1)=-\frac{2}{3}\psi^\prime(1)
-\frac{L^4}{18r_H^2}\phi^\prime(1)^2\psi(1) \; .
\end{eqnarray}
Using (\ref{regularity}) to eliminate $\psi^\prime$, we find
\begin{eqnarray}
\psi(z)=\frac{1}{3}\psi(1)+\frac{2}{3}\psi(1)z
-\frac{2}{9}\left(1+\frac{L^4}{8r_H^2}\phi^{\prime}(1)^2
\right)\psi(1)(1-z)^2
+\cdots
\label{psi:horizon}
\end{eqnarray}

\subsection{Solution near the asymptotic AdS region: $z=0$}
We expand $\phi$ and $\psi$, making use of asymptotic solutions 
(\ref{boundary}), as

\begin{eqnarray}
\phi(z)&=&\mu-qz+\frac{1}{2}\phi^{\prime\prime}(0)z^2+\cdots\\
\psi(z)&=&C_2z^2+\cdots
\end{eqnarray}
where we have used $C_1=0$.

Then the 2$^{\rm nd}$ derivative of $\phi$ is given by
\begin{eqnarray}
\phi^{\prime\prime}~\big|_{z=0}
=\frac{2L^2\psi^2}{z^2(1-z^3)}\phi~\bigg|_{z=0} = 0
\end{eqnarray}
and we get simply
\begin{eqnarray}
\phi(z)=\mu-qz\,,\hspace{1cm}\psi(z)=C_2z^2
\label{sol:infty}
\end{eqnarray}
where $qr_H$ is fixed.

\subsection{Matching and Phase Transition}

As before, we connect the solutions (\ref{phi:horizon}), (\ref{psi:horizon})
and (\ref{sol:infty}) at $z=\frac{1}{2}$. In order to connect those solutions smoothly, we require the following 4 conditions:
\begin{eqnarray}
&&\mu-\frac{1}{2}q=\frac{1}{2}b+\frac{L^2}{12}a^2b \label{c:phi}\\
&&-q=-b-\frac{L^2}{3}a^2b  \label{c:dphi}\\
&&\frac{1}{4}C_2=\frac{11}{18}a-\frac{L^4}{144r_H^2}ab^2 \label{c:psi}\\
&&C_2=\frac{8}{9}a+\frac{L^4}{36r_H^2}ab^2 \label{c:dpsi}
\end{eqnarray}
where $\psi(1)\equiv a$ and $-\phi^\prime(1)\equiv b$, 
with $(a,b>0)$ as before.
Eliminating $a^2 b$ from (\ref{c:phi}) and (\ref{c:dphi}) gives
\begin{eqnarray}
\mu = \frac{3}{4} q + \frac{1}{4} b \ . \label{relation:1}
\end{eqnarray}
From (\ref{c:phi}) and (\ref{c:dphi}), we can also deduce
\begin{eqnarray}
a^2 = \frac{12}{L^2 b} \left(q - \mu \right)  \ . \label{relation:2}
\end{eqnarray}
The above relation alludes to phase transitions, namely, given $q$,
$\mu$ has a maximum value when we assume the non-trivial solution $a\neq 0$.
Substituting the relation (\ref{relation:1}) into (\ref{relation:2}),
we have
\begin{eqnarray}
a = \frac{\sqrt{3}}{L} \sqrt{\frac{q}{b}} \sqrt{1-\frac{b}{q}} \ . 
\end{eqnarray}
To relate this result to the expectation value of the dimension 2 operator
$\langle{\cal O}_2\rangle = \sqrt{2} C_2 r_H^2/L^3$, 
we eliminate $ab^2$ from (\ref{c:psi}) and (\ref{c:dpsi})
to obtain
\begin{eqnarray}
C_2 = \frac{5}{3} a \ .
\end{eqnarray}
Similarly, eliminating $C_2$ from (\ref{c:psi}) and (\ref{c:dpsi}) gives
\begin{eqnarray}
a \left( b^2 -28 \frac{r_H^2}{L^4} \right) =0 \ ,
\end{eqnarray}
which determines $b= 2\sqrt{7} r_H/L^2$ provided $a\neq 0$.

Now we are in a position to reveal the phase transition phenomenon
in this simple system. Noting the relation $\rho=q~r_H$, and using
the Hawking temperature: $T=\frac{3r_H}{4\pi L^2}$, 
$\langle{\cal O}_2\rangle$ 
can be expressed by
\begin{eqnarray}
\langle{\cal O}_{2}\rangle
=\frac{80\pi^2 }{9}\sqrt{\frac{2}{3}}
T_c T\sqrt{1+\frac{T}{T_c}}\sqrt{1-\frac{T}{T_c}}
\end{eqnarray}
where $T_c$ is defined  as
\begin{eqnarray}
T_c=\frac{3\sqrt{\rho}}{4\pi L\sqrt{2\sqrt{7}}}
\label{Tc}
\end{eqnarray}
We see that $\langle{\cal O}_{2}\rangle$ is zero at $T=T_c$, 
which is a critical point, and
condensation occurs at $T<T_c$. The mean field theory result 
$\langle{\cal O}_{2}\rangle \propto(1-T/T_c)^{1/2}$
is also recovered. 
The value, (\ref{Tc}), of $T_c$ is evaluated as $0.103\sqrt{\rho}/L$. 
Comparing with the numerical result $0.118\sqrt{\rho}/L$ in \cite{HHH1},
we find our analytic approximation is quantitatively good.
Also the coefficient of $(1-T/T_c)^{1/2}$ as $T\rightarrow T_c$ is now 
$101T_c^2$, while the numerical result is $144T_c^2$ \cite{HHH1},
which means this approximation seems good.

One may wonder what happens if we change $z_m$. 
If we connect the solutions at $z_m\,\,(0<z_m<1)$, the result is
\begin{eqnarray}
\langle {\cal O}_{2} \rangle
=\frac{16\pi^2}{9}\frac{2+z_m}{3z_m}\sqrt{\frac{3}{1-z_m}}
T_cT\sqrt{1+\frac{T}{T_c}}\sqrt{1-\frac{T}{T_c}}
\end{eqnarray}
where
\begin{eqnarray}
T_c=\frac{3}{4\pi L}\sqrt{\frac{\rho}{\tilde b}}\,,\hspace{1cm}
{\tilde b}=\sqrt{\frac{4(1+5z_m)}{1-z_m}}
\end{eqnarray}
In order to get the same value $T_c=0.118\sqrt{\rho}/L$ as \cite{HHH1}, 
we need to choose $z_m=0.34$. For this value,
the coefficient of $(1-T/T_c)^{1/2}$ as $T\rightarrow T_c$ 
becomes $121 T_c^2$.
Thus, a numerically better approximation is possible, however, the
choice of $z_m$ does not give a big qualitative difference.

\end{document}